# A Novel High-Pressure Monoclinic Metallic Phase of $V_2O_3$

Yang Ding [1], Cheng-Chien Chen [1], Qiaoshi Zeng [2], Heung-Sik Kim [3], Myung Joon Han [3], Mahalingam Balasubramanian [1], Robert Gordon [1,4], Fangfei Li [5,6], Ligang Bai [7], Dimitry Popov [8], Steve M. Heald [1], Thomas Gog [1], Ho-kwang Mao [5,8,9,10], and Michel van Veenendaal [1,11]

[1] *Advanced Photon Source, Argonne National Laboratory, Argonne, Illinois 60439, USA*

[2] *Geological and Environmental Sciences, Stanford University, Stanford, California 94305, USA*

[3] *Department of Physics, Korea Advanced Institute of Science and Technology, Daejeon 305-701, Republic of Korea*

[4] *PNCSRF, APS Sector 20, Argonne, Argonne, Illinois 60439, USA*

[5] *HPSynC, Geophysical Laboratory, Carnegie Institution of Washington, Argonne, Illinois 60439, USA*

[6] *State Key Lab of Superhard Materials, Jilin University, Changchun 130012, China*

[7] *HiPSEC and Department of Physics, University of Nevada Las Vegas, Las Vegas, Nevada, 89154, USA*

[8] *HPCAT, Geophysical Laboratory, Carnegie Institution of Washington, Argonne, Illinois 60439, USA*

[9] *Geophysical Laboratory, Carnegie Institution of Washington, Washington, DC 20015, USA*

[10] *Center for High Pressure Science and Technology Advanced Research, 1690 Cailun Rd, Pudong, Shanghai 201203, P.R. China*

*and*

[11] *Department of Physics, Northern Illinois University, De Kalb, Illinois 60115, USA*

**Vanadium sesquioxide, $V_2O_3$, is a prototypical metal-to-insulator system where, in temperature-dependent studies, the transition always coincides with a corundum-to-monoclinic structural transition. As a function of pressure, $V_2O_3$ follows the expected behavior of increased metallicity due to a larger bandwidth for pressures up to 12.5 GPa. Surprisingly, for higher pressures when the structure becomes unstable, the resistance starts to increase. Around 32.5 GPa at 300 K, we observe a novel pressure-induced corundum-to-monoclinic transition between two metallic phases, showing that the structural phase transition can be decoupled from the metal-insulator transition. Using X-ray Raman scattering, we find that screening effects, which are strong in the corundum phase, become weakened at high pressures. Theoretical calculations indicate that this can be related to a decrease in coherent quasiparticle strength, suggesting that the high-pressure phase is likely a critical correlated metal, on the verge of Mott-insulating behavior.**

Metal-insulator transition, especially in strongly correlated materials, has been one of the oldest but still far from fully-understood topics in condensed matter physics [1, 2]. As a paradigmatic metal-to-insulator system, $V_2O_3$ has been the focus of six decades of extensive studies [3-12], but the underlying metal-to-insulator mechanism remains elusive. Various mechanisms have been proposed to account for the simultaneous electronic and structural transitions [13-29], but disentangling the contributions of the coupled spin, charge, and lattice variables remains a difficult task.

Phase transitions in $V_2O_3$ can be induced by changes in temperature, chemical composition, and external pressure [1]. At ambient conditions, $V_2O_3$ is a paramagnetic metal with the corundum structure (space group $R\bar{3}c$), where the vanadium atoms form V-V pairs along the crystal $c$-axis and honeycomb lattices in the $ab$-plane. Upon lowering the temperature below ~150 K, the paramagnetic $V_2O_3$ metal becomes an antiferromagnetic insulator, with a concomitant change of the crystal structure to monoclinic (space group $I2/a$) [4]. Doping with titanium ($V_{2-x}Ti_xO_3$) at low temperatures results in a transition from a monoclinic antiferromagnetic insulator to a corundum-structured paramagnetic metal. Until this study, all corundum-to-monoclinic structural transitions were accompanied by a metal-to-insulator transition. On the other hand, a paramagnetic-metal to paramagnetic-insulator transition can occur at high temperature in Cr-doped system ($V_{2-x}Cr_xO_3$) without a change in crystal symmetry. This leads to the important question of the relative importance of electron correlations (Mott physics) and electron-lattice coupling (Peierls distortion) in the metal-insulator transition.

In this Letter, we identify a new pressure-induced monoclinic metallic $V_2O_3$ phase using X-ray diffraction, electric resistance, and X-ray Raman experiments for pressures up to 49.7 GPa. As pressure has different effects on spin, orbital, charge, and lattice [30], it could potentially decouple these intertwined variables in the metal-insulator transitions [31]. For low pressures, applying pressure to $V_2O_3$ has a similar effect as chemical pressure by $Ti^{3+}$ doping, making the system more metallic due to an increase in bandwidth. However, we find that this analogy breaks down in the high-pressure regime: $V_2O_3$ becomes less metallic and its corundum structure is unstable above 13.5 GPa. The crystal structure eventually becomes monoclinic around 32.5 GPa at 300 K. This pressure-induced structural transition occurs without a concurrent metal-to-insulator transition, suggesting that lattice distortions cannot drive the system insulating if the electron correlations do not reach the critical value necessary to open an insulating gap.

Additionally, we study the pressure evolution of the electronic structure with X-ray Raman scattering at an X-ray energy loss range corresponding to the vanadium $M_{2,3}$-edge. In this technique, an electron from the shallow $3p$ core levels undergoes a sudden excitation into the unoccupied $3d$ valence states. The screening dynamics following this perturbation provides valuable information on the charge excitations. The ambient condition and low-pressure spectra are dominated by a single peak, which we describe theoretically using models based on exact diagonalization [32, 33]. To explain the spectral line shape, low-energy screening channels associated with coherent quasiparticle bands near the Fermi level $E_F$ are necessary. These quasiparticles (whose strength is the integrated intensity of the density of states close to $E_F$) behave essentially as electrons but with strongly renormalized properties, such as a larger effective mass, due to many-body interactions. Materials near a metal-insulator transition have significant incoherent quasiparticle spectral weight far away from $E_F$. This behavior has been predicted for correlated metals by techniques using combined local density approximation and dynamical mean field theory (LDA+DMFT) [17-19, 21, 22, 34]. In the high-pressure monoclinic phase, the X-ray Raman spectra are broadened by multiplet structures resembling closely those in the insulating compound. This reveals that the system, while still metallic (as indicated by the resistance), is unable to screen the Coulomb interactions resulting from the excitation created by the X-ray Raman process. The weakened screening can be attributed to a decreased strength of the coherent quasiparticle peaks near $E_F$ due to enhanced electronic correlations.

We first examine the effect of pressure on the electrical transport properties of $V_2O_3$ (powder samples purchased from Sigma-Aldrich with a purity of 99.99%). The detailed description of experiments can be found in "Supplemental Material". Figure 1 shows the results of isothermal resistance measurements under external pressure. In the low-pressure region (below 12.5 GPa), the resistance of $V_2O_3$ quickly drops from 6.4 ohm at 1.0 bar to 0.36 ohm at 12.5 GPa, which is consistent with previous studies [5, 6, 35, 36]. However, above 12.5 GPa the resistance starts to increase slowly with pressure and at 33.5 GPa reaches a value of 0.42 ohm, which is 17% more than that at 12.5 GPa. Above 33.5 GPa, the resistance increases even faster and the 0.52 ohm resistance at 47.5 GPa is roughly 52% more than the 12.5 GPa value. This shows that pressure makes $V_2O_3$ more metallic only up to 12.5 GPa. The unexpected result in the high-pressure regime (above 12.5 GPa) has not been predicted by theory before. Resistance depends on both the sample resistivity and the sample size, yet the change (increase) of resistance caused by sample size shrinkage is about one third of the compressibility [37], which is less than 3% below 12.5 GPa and even smaller at higher pressures due to a smaller sample compressibility and the almost constant thickness of cubic boron nitride/epoxy composite gasket [38]. Therefore, the abnormal increase of the resistance above 12.5 GPa should mainly result from the accompanying change in the crystal symmetry upon applying pressure.

We now discuss the X-ray powder diffraction measurements in characterizing the $V_2O_3$ crystal structure under pressure. Figure 2 displays the diffraction patterns measured from 3.2 GPa to 49.7 GPa, showing that the $V_2O_3$ sample remains in a corundum structure below 32.5 GPa. Although no apparent symmetry change could be identified below 32.5 GPa, the diffraction peaks are substantially broadened above 9.5 GPa. Moreover, the pressure-dependent lattice parameter $c$ shows an anomalous slope change at 13.0 GPa (see Fig. S1(b) in "Supplemental Material"), and above this pressure the residual of lattice refinement also suddenly increases. These are signs indicating that local structure distortion occurs around 9-13 GPa and continues to develop with pressure. We regard these local structure changes as the precursors of the structural phase transition happening above 32.5 GPa, effects that are commonly observed in the metal-to-insulator transitions of $V_2O_3$ [10, 39, 40]. As mentioned above, the isothermal electric resistance also starts to increases around 13.0 GPa, implying an underlying correlation between the structure and transport properties.

Above the critical pressure 32.5 GPa, the (012) and (110) diffraction peaks are split in the diffraction patterns, indicating a long-range order symmetry breaking, and thereby a structural phase transition. The diffraction patterns at the highest pressure 49.7 GPa are best indexed by the monoclinic lattices *P2$_1$/c* (space group No. 14) and *I2/a* (space group No.15). The volume of the monoclinic unit cell at 35.6 GPa is 181.41 Å$^3$, and the volume of the hexagonal unit cell (with six $V_2O_3$ molecules) at 28.6 GPa is about 273.15 Å$^3$. Taken together, this suggests a ~45.4 Å$^3$ molecular volume of $V_2O_3$ at the critical pressure, and the monoclinic cell possibly contains four $V_2O_3$ molecules, if we assume no large discontinuity in density near the transition. This assumption is indeed consistent with our structural analysis. As the low-temperature antiferromagnetic insulator phase also has the *I2/a* lattice symmetry and a similar unit cell volume [4], we first try to apply this structure as the seeding structure in the structure refinement. While the simulation matches fairly well with the experiments, it fails to reproduce the peak splitting at ~ 7.2° (2$\theta$) (see inset of Fig. 3a).

This mismatch suggests a lower symmetry of the high-pressure structure. The problem is solved by using the *P2$_1$/c* symmetry, and the peak splitting is accounted for by two nonequivalent V sites in *P2$_1$/c*, instead of the one unique V site in *I2/a*. The final refined and optimized monoclinic *P2$_1$/c* structure is shown in Fig. 3(d), and the atoms' Wyckoff positions are V1 4e (0.3455(2), -0.0008(4), 0.3007(8)), V2 4e (0.1514(2), -0.0002(5), 0.7018(8), O1 4e (0.4048(2), 0.8385(8), 0.6482(2)), O2 4e

(0.0944(7), 0.8371(1), 0.3547(2)), and O3 4e (0.2506(9), 0.3042(1), 0.5016(9))), respectively. Unlike the low-temperature monoclinic phase where the $V_2O_3$ volume expands, in the high-pressure phase the volume is reduced by pressure [Fig. S1(e)]. However, the V-V bond length is slightly elongated in the high-pressure phase above the critical pressure 32.5 GPa [Fig. S1(f)]. The crystal becomes increasingly difficult to be compressed along the $c$-axis, and the volume reduction is largely related to compression in the $ab$-plane as well as in the vacant space between $V_2O_3$ molecules. We note that while the two monoclinic phases $P2_1/c$ and $I2/a$ belong to distinct symmetry groups, they are in fact very similar in several aspects (see Fig. 3(b) and 3(c)).

To further study the pressure evolution of the electronic structure, we performed X-ray Raman scattering measurements on powdered $V_2O_3$ to investigate the vanadium $3p$-$3d$ excitations (corresponding to the $M_{2,3}$-edge). Figure 4(a) displays the spectra measured using the LERIX spectrometer at momentum transfer $q = 8.9$ Å$^{-1}$. All the spectra below the critical pressure 32.5 GPa are dominated by a single peak. The spectra for ambient condition $V_2O_3$ metal and for vanadium foil also show a similar feature. On the other hand, the 38 GPa spectrum in the monoclinic phase is largely broadened with an asymmetric two-peak profile. This substantial spectral-weight transfer over several electronvolts indicates a fundamental physical change in the system.

X-ray Raman scattering involves the excitation of core electrons into unoccupied states, and its cross section at small momentum transfer is closely related to electric dipole transition in X-ray absorption. With increasing magnitude of momentum transfer, higher multiple terms would dominate the transition matrix elements [41-44]. In our experimental setup, the $M_{2,3}$-edge spectra are governed by electric octupole transition (see "Supplemental Material"). The octupole spectrum close to the atomic limit shows an asymmetric two-peak structure (Fig. 4(b) black line) and spreads out in energy due to strong $3p$-$3d$ exchange interactions. The intensity of these "atomic" features is a good indication of the material's capacity to screen the Coulomb interactions. Our observation of a drastic change in the X-ray Raman spectra thereby implies an accompanying change in metallicity and electron correlation effects across the structure transition. On the other hand, while a strong *ad hoc* reduction in the interactions parameters could lead to a smaller multiplet energy spread, the interaction and local ligand hybridization parameters are not expected to differ substantially even across a metal-insulator-transition [45]. Therefore, the observed phenomenon requires further explanation.

Here we resort to crystal-field atomic multiplet calculations with additional screening channels arising from coherent quasiparticle bands near $E_F$, a feature manifest in correlated metals [17-19, 21, 22]. Similar approaches have been employed to study $V_2O_3$ [32, 33] and other systems [46-48], and prominent quasiparticle bands have been observed in photoemission experiments [25, 49-51]. The calculation (detailed in "Supplemental Material") contains five vanadium 3$d$ orbitals in an $O_h$ crystal field environment, which splits the 3$d$ orbitals into the $e_g^\sigma$ and $t_{2g}$ levels. An additional trigonal field of $D_{3d}$ symmetry for the corundum structure is used to lift the three-fold $t_{2g}$ degeneracy, resulting in an $e_g^\pi$ doublet and a single $a_{1g}$ level. To simulate the effects of quasiparticle screening, three "empty" and three "occupied" levels are included to hybridize with the $e_g^\pi$ and $a_{1g}$ orbitals. In the calculations, the lowest energy state has either $e_g^\pi - e_g^\pi$ or $a_{1g} - e_g^\pi$ triplet characters, consistent with previous effective $S = 1$ description of $V_2O_3$ [52-54].

Figure 4(b) displays the vanadium $M_{2,3}$-edge spectra computed with varying strengths of quasiparticle screening. The heavily-screened spectrum (red curve) shows a predominant single-peak feature, resulting from coherent quasiparticle levels strongly hybridized with the $e_g^\pi$ and $a_{1g}$ orbitals. Once the screening effect weakens (by shifting the quasiparticle band energy away from $E_F$), the spectrum is dominated by multiplet structures with an asymmetric two-peak profile (Fig. 4(b) black curve). We note that removing one of the screening channels already suffices to render a multiplet-dominated spectrum. Orbital-selective suppression of coherent quasiparticles has been suggested theoretically by Laad *et al.* [21, 22] where in the temperature-induced metal-to-insulator transition the quasiparticle bands associated with $e_g^\pi$ vanish near $E_F$, while those of $a_{1g}$ character remain more intact. Measurements such as X-ray absorption dichroism can further elucidate the orbital compositions of the pressure-induced monoclinic crystal.

In summary, our pressure-dependent room-temperature electronic and structural measurements on $V_2O_3$ show the presence of a high-pressure metallic monoclinic phase. At low pressures, this phase is preceded by a conventional corundum phase, showing a typical decrease in resistance due to an increase in bandwidth [5, 6, 35, 36]. However, around 12.5 GPa, the corundum structure becomes unstable and the resistance increases. The $M_{2,3}$-edge X-ray Raman spectra, on the other hand, do not exhibit any appreciable difference up to the critical pressure of 32.5 GPa. This indicates that the low-pressure corundum structured $V_2O_3$ remains a correlated metal with substantial coherent quasiparticle bands that allow the screening of the multiplet interactions. Therefore, the gradual increase in resistance between 12.5-32.5 GPa does not originate from a change in electron correlations, but rather from the underlying lattice distortion which suppresses charge fluctuations [28, 29]. However, across the critical pressure of 32.5 GPa, the electronic structure changes drastically with a spectral redistribution in the X-ray Raman spectra over several electronvolts, leading to a spectral line shape in resemblance to that of the insulating system. This clearly indicates that the electron correlations are enhanced in the high-pressure monoclinic phase, which directly explains the sudden increase in the slope of resistance versus pressure after the structural transition. However, the increase is below the critical value to open an insulating gap between the $e_g^\pi$ and $a_{1g}$ levels as evidenced by the metallic resistivity. Recent calculations by Poteryaev *et al.* [26, 27] suggest that the effective $e_g^\pi$ -$a_{1g}$ energy splitting $\Delta_{eff} = \Delta_{bare} + Re \sum a_{1g} - \sum e_g^\pi$ involves contributions from both the lattice and electron correlations, where $\Delta_{bare}$ is the bare crystal-field splitting, and a self-energy contribution due to electron interactions $Re \sum a_{1g} - \sum e_g^\pi$. Accordingly, a gap opens when $\Delta_{bare}$ or $Re \sum a_{1g} - \sum e_g^\pi$ increases. The X-ray Raman data clearly indicate an increase in the latter. However, in the high-pressure phase, the bare bandwidth increases due to a volume reduction, which together with the lowered monoclinic symmetry promotes hybridization and mixes the $e_g^\pi$ and $a_{1g}$ characters, thereby effectively decreasing $\Delta_{bare}$ [26, 27]. Therefore, the decrease in $\Delta_{bare}$ can compensate the change in self energy $Re \sum a_{1g} - \sum e_g^\pi$ resulting in an overall energy splitting $\Delta_{eff}$ too small to open a gap. Therefore, the high-pressure monoclinic $V_2O_3$ is likely a critical correlated metal, which is just on the verge of becoming a Mott insulator. Our discovery of this new phase could be regarded as evidence that the lattice degrees of freedom and electron correlations both contribute to the metal-to-insulator transitions in $V_2O_3$. At high pressures, electron-lattice coupling and electron correlation could compete for the occurrence of metal-to-insulator transitions and applying pressure can effectively decouple the metal-insulator transition from the structural phase transition and disentangle the coupled degrees of freedom. Future theoretical work and experiments comparing the high-pressure and low-temperature monoclinic phases could shed further light onto this problem.

**Figures and legends**

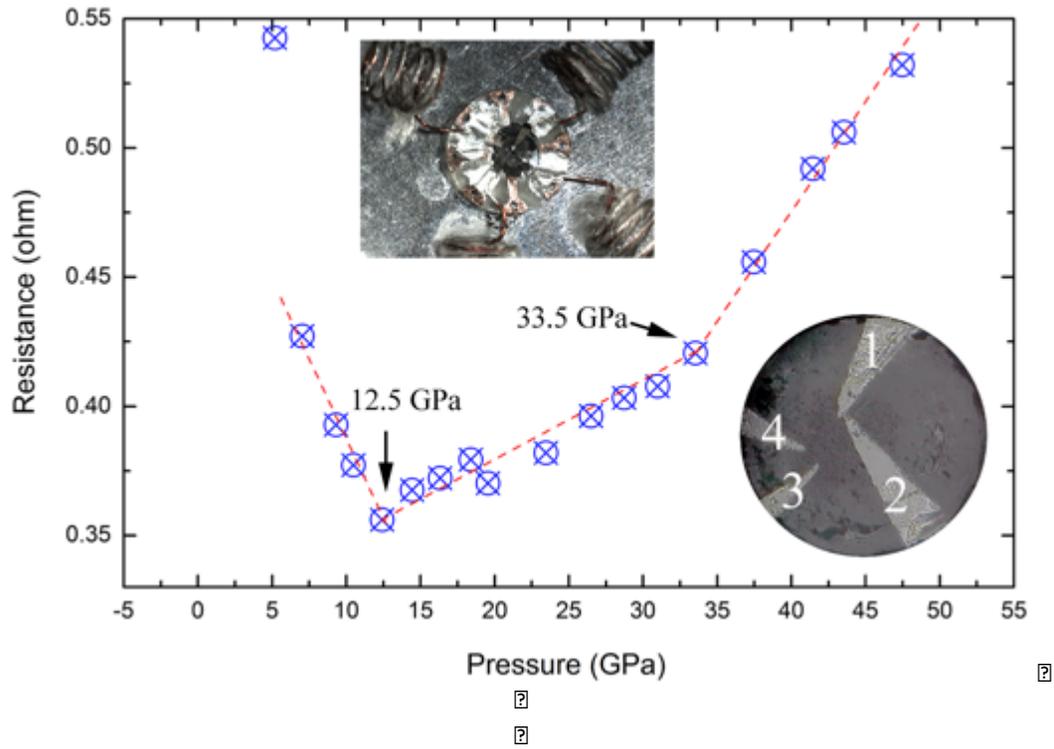

FIG. 1. Isothermal electric resistance of $V_2O_3$ under external pressure at 300 K. The top inset shows the four electric lines outside the gasket. The bottom right inset shows the four thin Pt probes attached to the sample, which are compressed by a pair of diamond anvils with culets of 400 $\mu m$. Pressure only makes the system more metallic up to 12.5 GPa, above which the electric resistance increases with pressure.

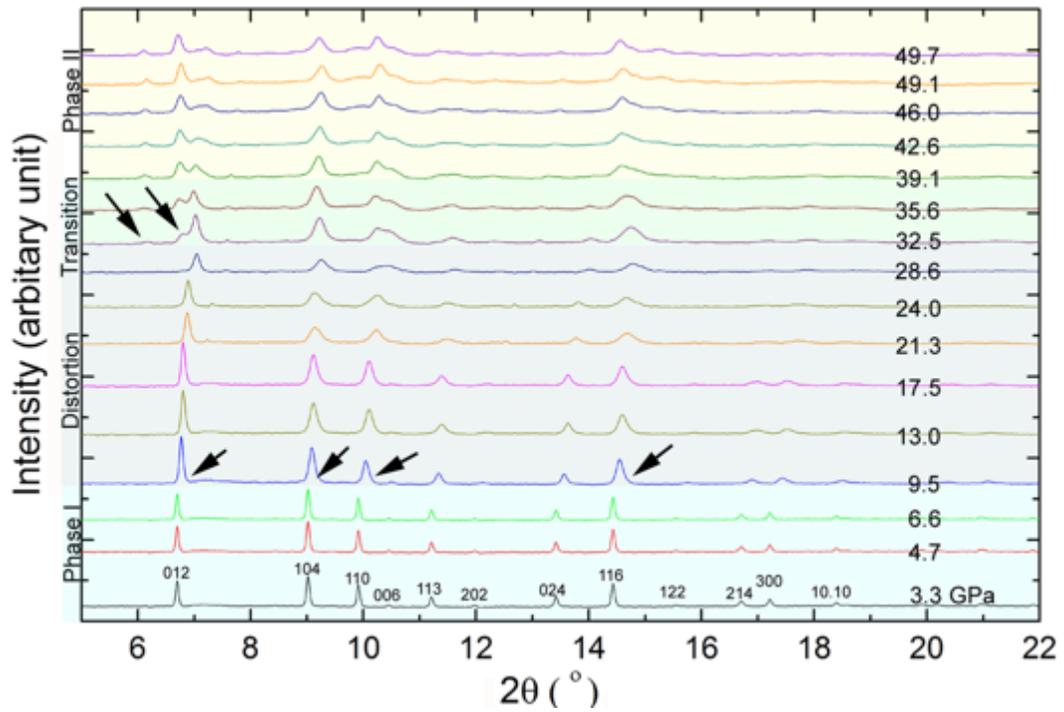

FIG. 2. The background-subtracted diffraction patterns of $V_2O_3$ under external pressure. The colored backgrounds represent the changes in the diffraction patterns. The patterns show significant peak broadening and lattice distortion above 9.5 GPa. Above 32.5 GPa the (012) and (110) peak splitting are clearly resolved, indicating a structural phase transition.

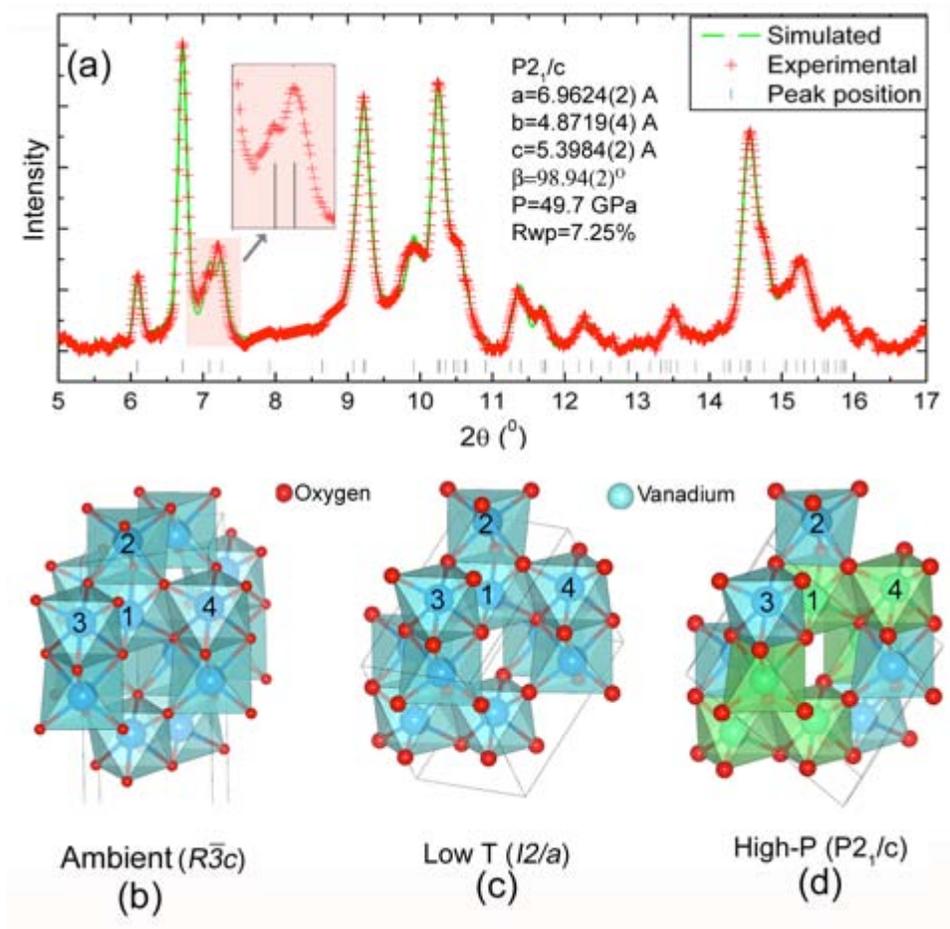

FIG. 3. (a) The structure refinement results, where the peak splitting at ~ 7.2° (2θ) is accounted for by the two nonequivalent V atoms in the high-pressure monoclinic phase. (b) Structure of the ambient condition corundum phase ($R\bar{3}c$). (c) Structure of the low-temperature monoclinic phase (*I2/a*). (d) Structure of the high-pressure monoclinic phase (*P2₁/c*); the blue and green colors represent nonequivalent V atoms. In (b)-(d), the thin black lines represent the unit cells, and the numbers are labels for the vanadium atoms.

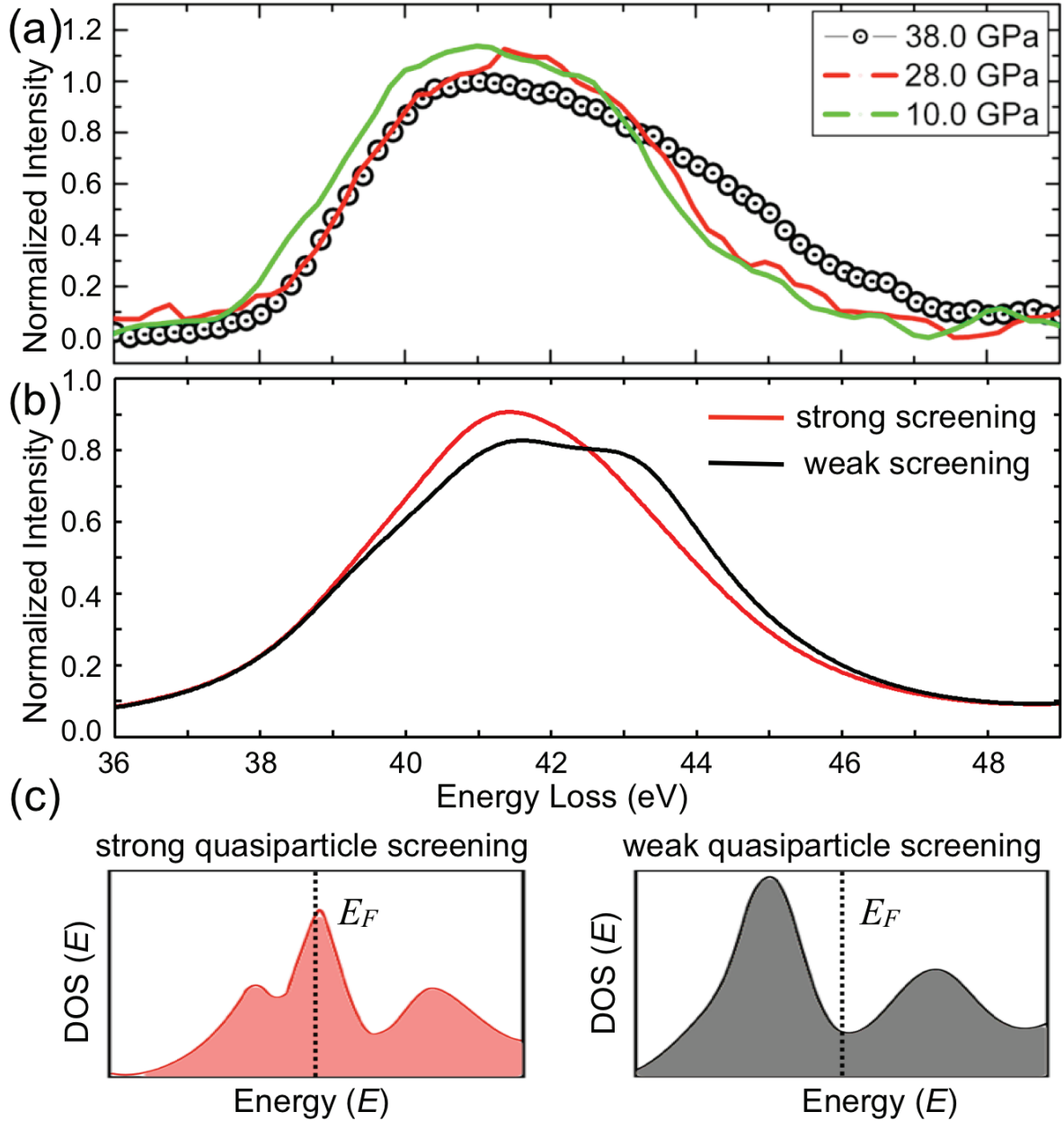

FIG. 4. The vanadium $M_{2,3}$-edge spectra from (a) experiments and (b) Theoretical calculations with varying strengths of quasiparticle screening effects. The low-pressure spectra show a predominant single-peak feature, while across the critical pressure 32.5 GPa the spectra are broadened by an asymmetric two-peak profile with multiplet structures. (c) Schematic plots showing the density of states (DOS) and the intensities of coherent quasiparticle bands near $E_F$ in the calculations.

We thank Dr. S. Tkachev for helium gas loading in the diffraction experiments. This research was supported in part by EFree, an Energy Frontier Research Center funded by the US Department of Energy (DOE), Office of Science, and Office of Basic Energy Sciences (BES) under Award No. DESC0001057. The diffraction measurements were carried out at HPCAT sector 16, and the X-ray Raman scattering measurements were carried out using the LERIX endstation at sector 20 (PNC/XSD). HP-CAT operations were supported by DOE-NNSA under Award No. DE-NA0001974, and DOE-BES under Award No. DE-FG02-99ER45775, with partial instrumentation funding by NSF. PNC/XSD facilities, and research at these facilities were supported by DOE-BES, a Major Resources Support grant from NSERC (Canada), the University of Washington, the Canadian Light Source and the Advanced Photon Source. Use of the Advanced Photon Source, an Office of Science User Facility operated for the U.S. Department of Energy (DOE) Office of Science by Argonne National Laboratory, was supported by the U.S. DOE under Contract No. DE-AC02-06CH11357. C.C.C. was supported by the Aneesur Rahman Postdoctoral Fellowship at ANL. H.K. Mao was supported as part of the EFree, an Energy Frontier Research Center funded by the U.S. Department of Energy, Office of Science, Office of Basic Energy Sciences under Award Number DE-SC0001057. M.v.V was supported by DOE-BES under Award No. DE-FG02-03ER46097 and the NIU Institute for Nanoscience, Engineering and Technology. The simulations were performed at NERSC - supported by DOE under Contract No. DE-AC02-05CH11231.